\documentstyle[prl,aps,eqsecnum]{revtex}
\input epsf

\begin{document}

\draft 

\twocolumn[\hsize\textwidth\columnwidth\hsize\csname
@twocolumnfalse\endcsname 

\title{Crossover to Potential Energy Landscape
Dominated Dynamics \\in a Model Glass-forming Liquid
}

\author{Thomas B. Schr\o der$^{1,2}$,
Srikanth Sastry$^{1,3}$, 
Jeppe C. Dyre$^2$, and 
Sharon C. Glotzer$^{1,4}$}

\address{$^1$Center for Theoretical and
Computational Materials Science, National Institute of Standards and
Technology, \\  Gaithersburg, Maryland, USA 20899}

\address{$^2$ Department of Mathematics and Physics (IMFUFA),
Roskilde University, PO Box 260,  DK-4000 Roskilde, Denmark.}	

\address{$^3$ Jawaharlal Nehru Centre for Advanced 
Scientific Research  Jakkur Campus, Bangalore 560064, INDIA
}	

\address{$^4$Polymers Division, National Institute of Standards and
Technology, Gaithersburg, Maryland, USA 20899}
\date{\today}

\maketitle

\begin{abstract}
An equilibrated model glass-forming liquid is studied 
by mapping successive configurations produced by  molecular dynamics 
simulation onto  a time series of inherent 
structures (local minima in the potential energy).
Using this ``inherent dynamics'' approach we find direct numerical 
evidence for the long held view that below a crossover temperature, 
$T_x$, the liquid's dynamics can be separated into (i) vibrations 
around  inherent structures and (ii) transitions between inherent 
structures (M. Goldstein, J. Chem. Phys. {\bf 51}, 3728 (1969)), 
i.e., the dynamics become ``dominated'' by the potential energy 
landscape.
In agreement with previous proposals, we find that $T_x$ is within 
the vicinity of the mode-coupling critical temperature $T_c$. 
We further find that at the lowest temperature simulated 
(close to $T_x$), transitions between inherent structures involve 
cooperative, string like rearrangements of groups of particles
moving distances substantially smaller than the average  interparticle 
distance.
\end{abstract} 

]

\narrowtext 
\section{Introduction}

Dynamical behavior of many physical and biological systems
\cite{Stillinger95,Angell95,confsub,foldfunnel} can be considered in
terms of the transient localization of the system in basins of
potential energy, and transitions between basins.  In particular, this
approach has received much attention in studies of slow dynamics and
the glass transition in supercooled liquids. Here, the strong
temperature dependence of transport properties such as the diffusion 
coefficient and
viscosity, and the possible existence of a thermodynamic transition
underlying the laboratory glass transition, have been sought to be
understood in terms of the properties of the liquid's potential energy
(or free energy) surface, or ``landscape'' as it is commonly called
\cite{Stillinger95,Angell95,Goldstein69,stillinger83,Angell88,FHS88,speedy,chandan,Sciortino,heuer,sastry,Schulz,ruocco,Sciortino99}.

For a system composed of $N$ atoms, the potential energy 
surface is simply the system's potential energy plotted as a 
function of the $3N$ particle coordinates in a $3N+1$ 
dimensional space \cite{Goldstein69}.
The potential energy surface contains a large number of local
minima, termed ``inherent structures'' by Stillinger and Weber
\cite{stillinger83}. Each inherent structure is surrounded by a
``basin'', which is defined such that a local minimization of the
potential energy maps any point in the basin to the inherent structure
contained within it.  The time evolution of a liquid may be viewed as
the motion of a point on the potential energy surface, and
thus as a succession of transitions from one basin to another.  These
transitions are expected to occur differently as the temperature $T$ is
varied.  In particular, Goldstein argued \cite{Goldstein69} that below
a crossover  temperature, $T_x$, where the shear relaxation time is
 $ \sim 10^{-9}$ seconds, relaxation is governed by thermally
activated crossings of potential energy barriers. 
The presence of significant energy barriers
below $T_x$ suggests a clear separation of short-time (vibrational)
relaxation within potential energy basins
from long-time relaxation due to transitions between basins.

A complementary approach to the dynamics of supercooled
liquids is provided by the mode coupling theory  (MCT) \cite{mct}. 
The simplest (so-called ``ideal'') version of this theory predicts 
a power-law divergence of 
relaxation times and the inverse diffusion coefficient, at a critical 
temperature $T_c$. Although a power law provides a reasonable description
of the temperature dependence of these quantities above $T_c$ in both 
real and simulated systems, power law behavior breaks down for 
$T\approx T_c$,  i.e. the predicted singularity at $T_c$ is not observed. 
This deviation is attributed to the presence of ``hopping'' motion as 
a mechanism of relaxation, which is not included in ideal MCT \cite{mct}.
Consequently, $T_c$ is usually estimated by 
fitting a power law to a relaxation time, taking into account
that this fit is expected to break down close to (and below) $T_c$.

It was noted by Angell \cite{Angell88}  that 
experimentally it is often found that the shear relaxation 
time is on the order of $10^{-9}$ seconds at the estimated  $T_c$, 
leading to the argument that $T_x \approx T_c$ (See also
Ref.\cite{Sokolov98}).
The presence of a low temperature regime where
barrier crossings dominate the dynamics, and the correspondence of the
crossover to that regime with the mode coupling critical temperature 
$T_c$, has also been  discussed in the context of mean field theories 
of certain spin glass models \cite{kirk,parisi,franz}. 

The existence of a crossover temperature and corresponding separation
of the dynamics  can be directly tested with computer simulations, 
using the concept of inherent structures.  In
this paper, we map the dynamical evolution of an equilibrated model
liquid to a time series of inherent structures for a range of 
temperatures. In this way, we test the extent to
which short-time ``intra-basin'' relaxation is separable from
long-time ``inter-basin'' relaxation. Our results demonstrate that 
this separation becomes valid as the system is cooled, and we estimate 
the crossover temperature $T_x$ to be close to the 
estimated value of $T_c$.

\section{Inherent Dynamics}

\begin{figure}
\hbox to\hsize{\epsfxsize=1.0\hsize\hfil\epsfbox{
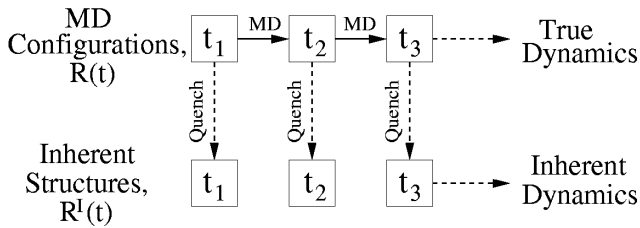}\hfil}
\caption{Schematic describing the principle of the ``inherent
dynamics'' approach. Successive configurations of the equilibrated
liquid ${\bf R}(t)$ are quenched to produce their corresponding
inherent structures ${\bf R}^I(t)$. Successive inherent structures
form a time series which we use to calculate the inherent self intermediate 
scattering function  $F_{s}^I({q}, t)$ (Eq. \ref{FsIDef}).  More 
generally, the inherent counterpart of any equilibrium quantity
may be calculated in this fashion.
}
\label{fig:Schematic}
\end{figure}

In this section we describe the details of our approach, which is
sketched in Fig. \ref{fig:Schematic}.
After equilibration at a given thermodynamic state point, 
a discrete time series of configurations,  ${\mathbf R}(t)$, 
is produced by standard molecular dynamics (MD) simulation.
Each of the configurations ${\mathbf R}(t)$ is then mapped to its
corresponding inherent structure, ${\mathbf R}^I(t)$, by locally 
minimizing the potential energy in configuration space.
We refer to this procedure as a ``quench''.
After quenching the configurations in ${\mathbf R}(t)$, we have 
two ``parallel'' time series of configurations,  
${\mathbf R}(t)$ and ${\mathbf R}^I(t)$.
The time series  ${\mathbf R}(t)$ defines the ``true dynamics'', which is
simply the usual (Newtonian) MD dynamics. In an analogous way, 
the time series ${\mathbf R}^I(t)$ defines the ``inherent dynamics''.
If a function quantifying some aspect of  the true dynamics 
is denoted by $f({\mathbf R}(t))$, then the corresponding 
function, $f({\mathbf R^I}(t))$, of the inherent dynamics  is calculated in 
exactly the same way, except using the  time series of inherent structures. 
For example, the self intermediate scattering function, 
$F_{s}({q}, t)$,
and the \emph{inherent} self intermediate scattering function, 
$F_{s}^I({q}, t)$, are  defined by 
\begin{eqnarray}
    F_{s}({q}, t) &\equiv& 
      \langle 
         \cos{{\mathbf q}\cdot ({\mathbf r}_j(t) - {\mathbf r}_j(0))} 
      \rangle \label{FsDef} \mbox{~,~~~ and}\\
    F_{s}^I({ q}, t) &\equiv& 
      \langle 
         \cos{{\mathbf q}\cdot ({\mathbf r}^I_j(t) - {\mathbf r}^I_j(0))} 
      \rangle \label{FsIDef}
\end{eqnarray}
where ${\mathbf r}^I_j(t)$ is the position of the $j$th particle 
in the inherent structure ${\mathbf R}^I(t)$ and $\langle ... \rangle$ 
denotes an average over $j$ and the time origin.

In this paper, we quantitatively compare $F_{s}({q}, t)$ and 
$F_{s}^I({q}, t)$ to test whether the dynamics of a binary 
Lennard-Jones mixture can be separated into vibrations around, and 
transitions between inherent structures. If so, then
$F_{s}^I({q}, t)$ describes the relaxation of the liquid 
as described by $F_{s}({q}, t)$, but with the effect of the 
vibrations removed. We show that this scenario becomes true below 
a crossover temperature, $T_x$, which is close to the lowest temperature
simulated in the present work.

\section{Results}

In the following we present results from molecular dynamics simulations 
of  a binary Lennard-Jones mixture in three dimensions, equilibrated at 
eight different temperatures.
The model used for the present simulations is described in 
Ref. \cite{SchroederDyre}. The system contains $251$ particles of type
A and $249$ particles of type B interacting via a binary Lennard-Jones
potential with parameters $\sigma_{BB}/\sigma_{AA} = 5/6$,
$\sigma_{AB} = (\sigma_{AA} + \sigma_{BB})/2$, and $\epsilon_{AA} =
\epsilon_{AB} = \epsilon_{BB}$.  The masses are given by $m_B/m_A =
1/2$.  The length of the sample is $L=7.28\sigma_{AA}$ and the
potential was cut and shifted at $2.5\sigma_{\alpha \beta}$.  All
quantities are reported in reduced units: $T$ in units of
$\epsilon_{AA}$, lengths in units of $\sigma_{AA}$ and time in units
of $\tau \equiv (m_{B} \sigma_{AA}^2/48\epsilon)^{1/2}$ 
(this was misprinted in \cite{SchroederDyre}). Adopting  
``Argon units'' leads to  $\sigma_{AA}=3.4 {\rm\AA}$, 
$\epsilon/k_B=120K$, and $\tau = 3\times10^{-13}$s.
The simulations were performed in the NVE ensemble using the leap-frog
algorithm with a timestep of $0.01\tau$, at constant reduced density, 
$\rho = 1.296$. The quenching was performed using the conjugate gradient
 method \cite{Press}.

We first briefly describe aspects of the true dynamics that demonstrate
a qualitative change occuring in the temperature range investigated.

\begin{figure}
\hbox to\hsize{\epsfxsize=1.0\hsize\hfil\epsfbox{
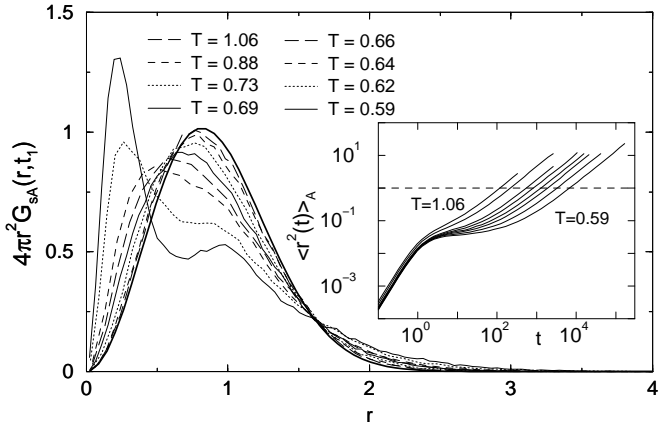}\hfil}
\caption{
Distribution of particle displacements for the A (large) particles, 
$4\pi r^2G_{sA}(r,t_1)$, where $t_1$ is defined by $\langle
r^2(t_1)\rangle_A=1$ (see inset).    
At high $T$ the Gaussian
approximation (thick curve) is reasonable, whereas at the lowest $T$ a
second peak is present, indicating single particle hopping. Inset:
Mean square displacement of the A particles, $\langle r^2(t)\rangle_A$.
 Similar behavior
is found for the B (small) particles.
}
\label{fig:VanHove}
\end{figure}

\begin{figure}
\hbox to\hsize{\epsfxsize=0.8\hsize\hfil\epsfbox{
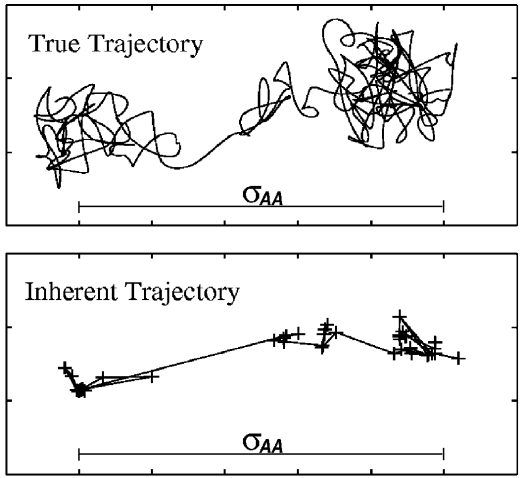}\hfil}
\vspace{.5cm}
\caption{
(a) Trajectory of a particle at $T=0.59$.
The elapsed time is $\Delta t= 160\tau$ (the typical ``vibration'' time is
$\approx 1\tau$).  At this temperature the dynamics is dominated by 
``hopping'': particles stay relatively
localized for many time steps and then move approximately one interparticle 
distance, where they again become localized.
(b) Applying the inherent dynamics approach to the trajectory above. 
The 1600 configurations used to generate the (true) trajectory in (a) were
quenched, and the positions of the particle in the resulting inherent 
structures are here plotted and connected by straight lines.
}
\label{fig:Trajectories}
\end{figure}

In Fig.~\ref{fig:VanHove} we show the quantity $4\pi r^2G_{sA}(r,t_1)$,
which is the distribution of  displacements \cite{Hansen} of particles of 
type A during the time
interval $t_1$. We define $t_1$ as the time where the mean square displacement
is unity, $\langle r^2(t_1)\rangle_A=1$. 
At all temperatures the dynamics become diffusive 
($\langle r^2(t)\rangle_A \propto t$) for $t \gtrsim t_1$ (see inset), 
i.e., $t_1$ marks the onset of diffusivity.
At the highest temperatures, $4\pi r^2G_{sA}(r,t_1)$ agrees well with the
Gaussian approximation [thick curve, $G_{sA}(r,t_1)\propto
\exp(-3r^2/2)$].  As $T$ is lowered, the distribution of particle 
displacements deviates from
the Gaussian approximation, and a shoulder develops at the average
interparticle distance ($r\approx 1.0$ in the adopted units), 
which at T=0.59 becomes a
well-defined second peak.  The second peak, observed also in other
model liquids at low temperatures, indicates \cite{Roux,Wahnstrom} single 
particle ``hopping'' (see Fig.~\ref{fig:Trajectories}a): 
particles stay relatively localized for a period of 
time (first peak), and then move approximately one interparticle distance, 
where they again become localized (second peak).
Thus we see from Fig. \ref{fig:VanHove} that as we approach our lowest 
simulated temperature $T=0.59$, there is a qualitative change from 
dynamics well described by a Gaussian distribution to dynamics dominated
by hopping processes.
 
In Fig.~\ref{fig:Trajectories}b the inherent dynamics approach is applied
to the true trajectory seen in Fig.~\ref{fig:Trajectories}a. The resulting
``inherent trajectory'' consists of the positions of the particle in 
1600 successive quenched configurations. 
The quenching procedure is seen to remove the vibrational motion from the 
true trajectory.  The inherent trajectory will be discussed in more 
detail in section \ref{sec:transitions}.

\begin{figure}
\hbox to\hsize{\epsfxsize=1.\hsize\hfil\epsfbox{
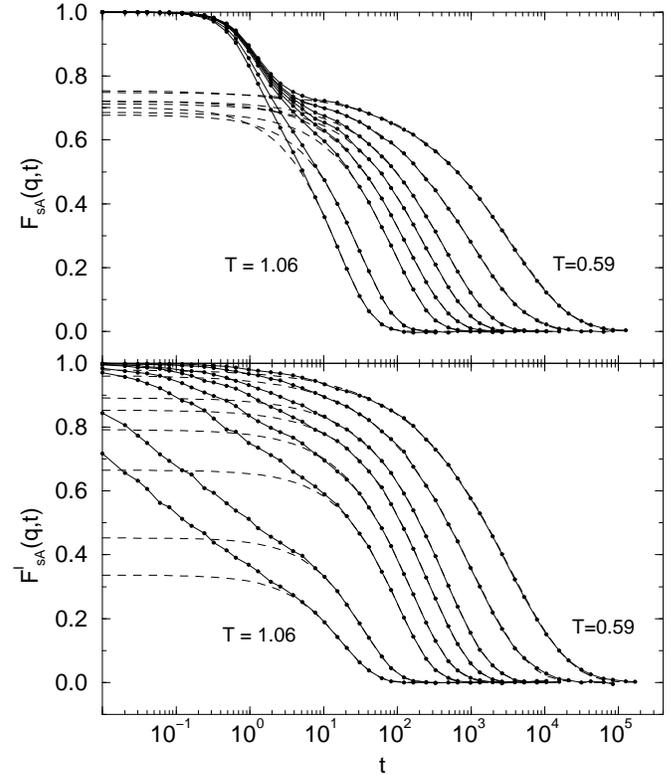}\hfil}
\caption{(a) $F_{sA}(q, t)$ plotted versus $t$ on log-scale for $q=7.5$ at 
the same temperatures as in figure \ref{fig:VanHove}.
Data points are connected by straight lines. 
Dashed lines are fits to $f(t) = f_c\exp(-(t/\tau_{\alpha})^{\beta})$.  (b)
$F_{sA}^I(q, t)$, otherwise as above. 
In both (a) and (b), the fitting was performed for $t>10$ for the 
two highest temperatures and for $t>30$ otherwise.
Similar behavior is found for the B particles.}
\label{fig:Intermediate}
\end{figure}
We now compare the true self intermediate scattering function, 
$F_{s}({q}, t)$, with its inherent counterpart $F_{s}^I({q}, t)$.
Fig.~\ref{fig:Intermediate}a shows the self intermediate
scattering function for the A particles, $F_{sA}(q, t)$, at $q=7.5$
corresponding to the position of the primary peak in the 
static structure factor for the A-A correlation. 
For each temperature $F_{s}({q}, t)$ was calculated from 
approximately 2000 configurations (depending on temperature).
As $T$ decreases, $F_{sA}(q, t)$ is found to display the typical 
two-step relaxation, where the short time
decay is  attributed to vibrational relaxation (or ``dephasing'',
see Ref. \cite{Sciortino}) of particles within cages formed by
neighboring particles \cite{Yonezawa94a,Kudchadkar95,Sciortino96}.  
The long time, or $\alpha$-relaxation is
separated from the short time regime by a plateau indicating transient
localization, or ``caging'' of particles, and is generally observed  
to follow a stretched exponential form.

The self part of the inherent 
intermediate scattering function for the A particles, $F_{sA}^I(q,
t)$ at q=7.5, is shown in Fig.~\ref{fig:Intermediate}b. This was calculated by 
quenching each configuration used in Fig.~\ref{fig:Intermediate}a, and
then applying the same data analysis program on the resulting time
series of inherent structures. As expected, the
plateau disappears in the inherent dynamics, as previously shown also for
the inherent mean-square displacement \cite{SchroederDyre}.  
At all $T$ we find that the long-time behavior of both $F_{sA}(q, t)$ and
$F_{sA}^I(q, t)$ is well described by stretched exponentials (dashed
lines). As a result, we can quantitatively compare 
the long time relaxation of $F_{sA}(q, t)$ and $F_{sA}^I(q, t)$,
by comparing the fitting parameters $\{ \tau_\alpha, \beta, f_c \}$
of the stretched exponentials  $f(t) = f_c\exp(-(t/\tau_{\alpha})^{\beta})$.

If the true dynamics can be separated into 
vibrations around and transitions between inherent structures, 
how do we expect the fitting parameters for
the inherent self intermediate scattering function,
$\{ \tau_\alpha^I, \beta^I, f_c^I \}$ to be related to  
the fitting parameters for the true  self intermediate scattering 
function, $\{ \tau_\alpha, \beta, f_c \}$? 
To answer this question, we assume that the initial relaxation in 
$F_{s}({ q}, t)$ is due to vibrations (as widely accepted 
\cite{Sciortino,Yonezawa94a,Kudchadkar95,Sciortino96}).
If this is the case, then we expect the quenching procedure to remove 
the initial 
relaxation (since it removes the vibrations), which means that  
$F_{s}^I({ q}, t)$ can be thought of as $F_{s}({ q}, t)$ with the initial 
relaxation removed\footnote{If vibration can be separated from transitions
between inherent structures  we may write for the x-displacement: 
$ \Delta x=\Delta x_{vib}+\Delta x_{inh}$ where the two terms are
 statistically uncorrelated. Thus, [using an exponential instead
 of cosine in Eqs. (2.1) and (2.2)] we find that the self intermediate
 scattering function is a {\em product} of a term relating to vibrations
 and one relating to transitions between inherent structures. At long
 times the former becomes time-independent, converging 
to the non-ergodicity parameter.}. 
This in turn means that $F_{s}^I({ q}, t)$ should be 
identical to the long time relaxation of $F_{s}({ q}, t)$, but rescaled 
to start at unity:  
$\{ \tau_\alpha^I, \beta^I, f_c^I \} = \{ \tau_\alpha, \beta, 1\}$.

\begin{figure}
  \hbox to\hsize{\epsfxsize=1.\hsize\hfil\epsfbox{
  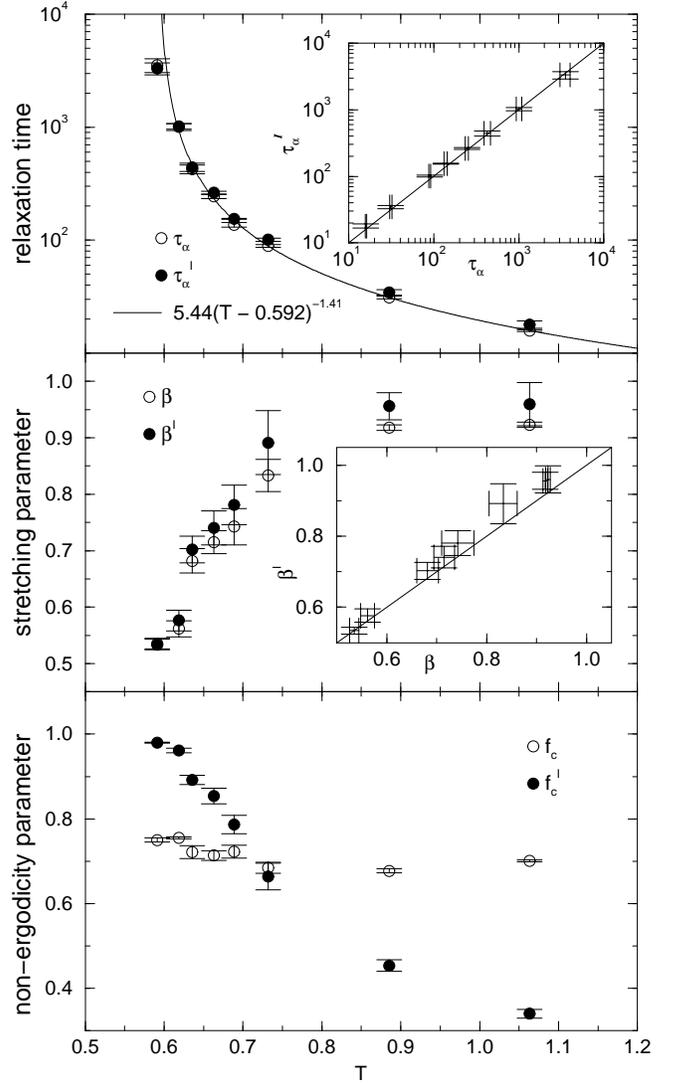}\hfil}
\caption{ Parameters describing the fit of $F_{sA}(q = 7.5, t)$ and
$F_{sA}^I(q = 7.5, t)$ to stretched exponentials from Fig.~3 (a) and
(b), respectively. (a) Relaxation times $\tau_{\alpha}$ and 
$\tau_{\alpha}^I$
vs. $T$. The solid line is a fit to 
$\tau_{\alpha} \propto (T - T_c)^{-\gamma}$
excluding the lowest $T$ in the fitting (see text). 
INSET: $\tau_{\alpha}^I$ vs. $\tau_{\alpha}$.  (b) Stretching
parameters $\beta$ and $\beta^I$ vs. $T$. INSET: $\beta^I$
vs. $\beta$.  (c) Non-ergodicity parameters $f_c$ and $f_c^I$
vs. $T$. Error bars are estimated from deviations between three
independent samples. Similar behavior is found for the B particles.
}
\label{fig:FsParameters}
\end{figure}

The fitting parameters used for fitting stretched exponentials to
$F_{sA}(q, t)$ (Fig.~\ref{fig:Intermediate}a) and $F_{sA}^I(q, t)$
(Fig.~\ref{fig:Intermediate}b)  are shown in
Fig.~\ref{fig:FsParameters}: (a) relaxation times, $\tau_{\alpha}$ and
$\tau_{\alpha}^I$, (b) stretching parameters, $\beta$ and $\beta^I$, and
(c) non-ergodicity parameters, $f_c$ and $f_c^I$.  We also show in
Fig.~\ref{fig:FsParameters}a the fit of the asymptotic mode coupling
prediction $\tau_{\alpha} \propto (T - T_c)^{-\gamma}$, from which we
find $T_c = 0.592 \pm 0.006$ and $\gamma = 1.41 \pm 0.07$.  The fitting
was done without the lowest temperature, where hopping is clearly
present in the system (see Fig.~\ref{fig:VanHove}), since this type of
particle motion is not included in the ideal mode coupling
theory. Excluding the \emph{two} lowest $T$ gives a fit which is
consistent with the one presented here; including all temperatures
gives a considerably worse fit.  Applying the same procedure to the
inverse diffusion coefficient, $D^{-1}(T)$, gives $T_c = 0.574 \pm 0.005$ and 
$\gamma = 1.40 \pm 0.09$ (data not shown). 

Also shown in Fig.~\ref{fig:FsParameters} as insets are
$\tau_{\alpha}^I$ vs. $\tau_{\alpha}$ and $\beta^I$ vs. $\beta$.  
Within the error bars we find that $\tau_{\alpha}$ and $\tau_{\alpha}^I$ 
are identical at all temperatures. At the highest temperatures
$\beta$ is poorly defined since there is no well-defined plateau 
in $F_{sA}(q, t)$. Consequently it is  difficult to compare $\beta$ and 
$\beta^I$ at high $T$, but we find that they become identical 
(within the error bars) at low $T$. Thus at low temperatures our results
confirms the expectation that the inherent dynamics is simply a
 coarse-graining of
the true dynamics, i.e., that
$\{ \tau_\alpha^I, \beta^I \} = \{ \tau_\alpha, \beta\}$.
On the other hand, the non-ergodicity parameters $f_c$ and $f_c^I$ 
(Fig.~\ref{fig:FsParameters}c) are
strikingly different.  While $f_c$ is roughly independent of $T$,
$f_c^I$ increases towards unity as 
$T$ approaches our lowest temperature. The fact that we  observe 
 a temperature dependence of $f_c^I$ approaching unity as 
$T$ approaches our lowest temperature $T=0.59$, leads us to conclude that this 
is close to the crossover temperature, $T_x$.
We note  that Goldstein's  estimate of shear relaxation
times at $T_x$ ($10^{-9} $ seconds) in our LJ units corresponds
to  $3\times 10^{3}$, which is the same order of magnitude as $\tau_\alpha$
in the temperature range where $f_c$ approaches unity.

Below $T_x$ the inherent dynamics can be thought of as the true 
dynamics with the effect of the vibration removed, as shown above.
How should the inherent dynamics be interpreted {\em above} $T_x$?
In Fig.~\ref{fig:Intermediate}b the short time relaxation of the 
inherent self intermediate scattering function at high temperatures is 
seen to be approximately logarithmic in time. This is an artificial 
relaxation introduced by 
applying the quenching procedure at a temperature where 
the dynamics is \emph{not} separated into vibrations around, 
and transitions between inherent structures, i.e. the  quenching procedure 
is doing more than simply removing the vibrations around inherent structures. 
Presumably the inherent dynamics above $T_x$
contains information about the underlying potential energy landscape.
At the present, however, we do not know how to interpret this, and we 
do not have an explanation as to why the (artificial) initial relaxation
appears to be  logarithmic at high temperatures.

We now proceed to discuss Angell's proposal, that $T_x \approx T_c$.  
We find that both estimated values for $T_c$ [$0.592\pm 0.005$ from 
$\tau_\alpha(T)$ and $0.574\pm 0.005$ from $D^{-1}(T)$] are in the 
temperature range where $f_c^I$  is approaching unity. 
We note that in the system investigated here two of the 
asymptotic predictions of the ideal mode coupling theory do not hold;
$\tau_\alpha$ and
$D^{-1}$ have different temperature dependence and we do not find
time-temperature super-position of the $\alpha$-part of  
the self intermediate scattering function. 
However, the argument given by Angell \cite{Angell88} (and Sokolov 
\cite{Sokolov98}) only relates to $T_c$ as the temperature where 
power-law fits to experimental data tend to break down, i.e. the 
``usage'' of MCT in this argument is similar to the way we have 
estimated $T_c$ in Fig. \ref{fig:FsParameters}a, 
and does not require, e.g., time-temperature superposition.

\section{Transitions between inherent structures}
\label{sec:transitions}

\begin{figure}
\hbox to\hsize{\epsfxsize=1.0\hsize\hfil\epsfbox{
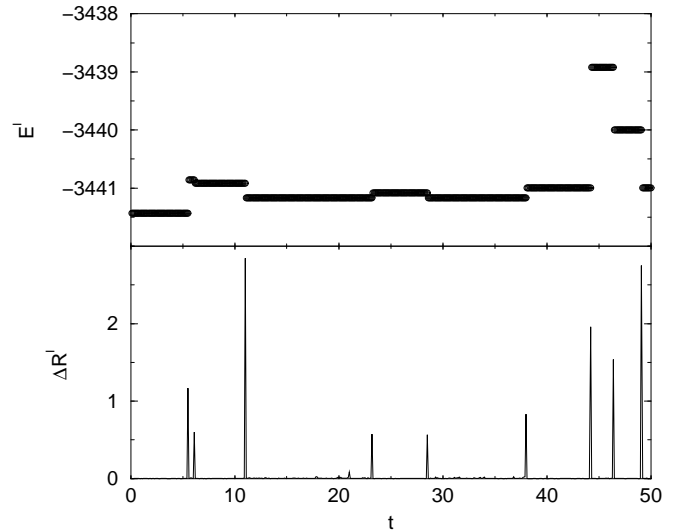}\hfil}\caption{
 Identifying transitions between inherent structures. (a) The inherent 
structure energy $E^I(t)$ vs. time.
(b) $\Delta  R^I$ (Eq. \ref{DeltaR}) vs. time. A transition between 
(the basin of attraction of) two inherent structures is indicated by a jump in 
 $E^I(t)$ and a corresponding peak in $\Delta R^I(t)$.
} 
\label{fig:IdentTrans}
\end{figure}

As shown in the previous section, separation of the dynamics into 
vibrations around and transitions between (the basin of attraction of)
inherent structures
becomes possible as $T$ approaches $T_x$, which is close to our
lowest simulated temperature T=0.59. At this temperature, it therefore
becomes meaningful  to examine the details of the  transitions between 
successive inherent structures. We identify such transitions 
by quenching the MD configurations every $0.1\tau$ (i.e. every 10
MD-steps) and looking for signatures of the system undergoing a 
transition from one inherent structure to another. We have considered
2 such signatures: i) We monitor the inherent structure energy $E^I(t)$ 
as a function of time, as shown in Fig. \ref{fig:IdentTrans}a. 
ii) We monitor the distance in configuration 
space $\Delta R^I(t)$ between two successive quenched
configurations \cite{Ohmine95} (Fig. \ref{fig:IdentTrans}b), where  

\begin{eqnarray}
   \Delta R^I(t) &\equiv& |{\mathbf R}^I(t+0.1) - {\mathbf R}^I(t)| \label{DeltaR}\\
   &=& \sqrt{\sum_{j=1}^{N}\left({\mathbf r}^I_j(t+0.1) - {\mathbf r}^I_j(t)\right)^2}.
\end{eqnarray}
Each jump in $E^I(t)$ corresponds to a peak in 
$\Delta R^I(t)$, indicating a transition to a new inherent structure.
In the (rare) event where a transition occurs between  two  inherent structures
with the same energy, $\Delta R^I(t)$ will still exhibit a peak even in the 
absence of a jump in $E^I(t)$, and  for this reason we  use  
$\Delta R^I(t)$ to identify transitions. The condition $\Delta R^I(t)>0.1$
was  found to be a sufficient threshold for this purpose. When evidence of a
transition was found in a time interval $\Delta t=0.1\tau$, this time interval 
was divided into 10 subintervals of $\Delta t=0.01\tau$ and the procedure 
described above was repeated. 

\begin{figure}
\hbox to\hsize{\epsfxsize=1.0\hsize\hfil\epsfbox{
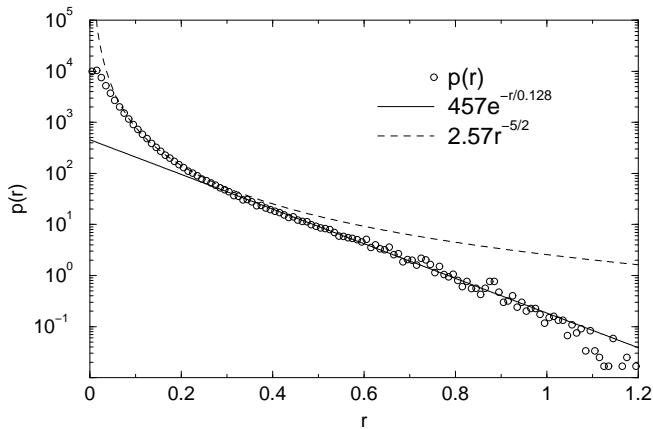}\hfil}
\caption{
Distribution of particle displacements during transitions between 
successive inherent structures at T=0.59. The integral of the distribution
is normalized to be the number of particles $N=500$.
Full curve is a fit to an exponential, for $0.3<r<1.0$.
The dotted curve is a fit to a power-law with (fixed) exponent $-5/2$
\protect\cite{Dyre99}, for $0.1 < r < 0.2$
} 
\label{fig:DisplDistr}
\end{figure}

For each transition, we monitor the difference between the particle 
positions in the two successive inherent structures. 
The distribution $p(r)$ of all such 
particle ``displacements'' averaged over the 12000 transitions 
we have identified is shown in Fig. \ref{fig:DisplDistr}. 
While many particles move only a small distance ($r<0.2$)
during a transition from one inherent structure to the next, 
a number of particles move farther, and in
particular, we find that the distribution for $r>0.2$ is to 
a good approximation exponential. 
The dotted curve is a fit to a power-law with exponent $-5/2$, 
which is a prediction from linear elasticity theory \cite{Dyre99}, 
describing the displacements of particles in the surroundings of
a local rearrangement ``event''. This  power-law fit does not look 
very convincing  by itself, but we note that the exponent was not 
treated as a fitting
parameter (i.e. only the prefactor was fitted), and the power-law {\em must}
break down for small displacements, since these correspond
to distances far away from the local event, and are thus not present  in
our relatively small sample. From the change in behavior of $p(r)$ at 
$r \approx 0.2$,
it is reasonable to think of particles with displacements larger than $0.2$
as those taking part in the local event, and the rest of the particles as
merely ``adjusting'' to the local event\footnote{Note however, that our 
data does
not imply what is cause and what is effect, or even if such a distinction
is meaningful.}. Using this 
definition it is found that on average approximately 10 particles 
participate in an event.

Fig. \ref{fig:DisplDistr} has two important consequences with regards
to points discussed earlier in this paper. 
The first point relates
to the single particle hopping indicated by the 
secondary peak in $4\pi r^2 G_s(r,t)$ (Fig. \ref{fig:VanHove}) 
at low temperatures. A common interpretation of the single particle 
hopping is that the jump of a  particle from one ``localized state'' 
(first peak) to the second localized state (secondary peak), 
corresponds to the transition of the system over an energy barrier 
from one inherent structure to the next. 
If such a transition typically occurs over a single energy 
barrier, i.e. without any new inherent structures between the two states, 
we would expect to find a preference for displacements of one average 
interparticle distance ($r\approx 1$) in Fig. \ref{fig:DisplDistr}. 
That this is {\em not} the case demonstrates that the hopping indicated by the 
secondary peak in $4\pi r^2 G_s(r,t)$ 
at low temperatures is {\em not} due to transitions over single 
energy barriers. Instead, as seen  in 
the inherent trajectory in Fig. \ref{fig:Trajectories}, the jump  
occurs via a number of ``intermediate'' inherent structures.

The second important consequence of Fig. \ref{fig:DisplDistr} is that
particles in the surroundings of a local event are displaced 
by small distances. This kind of motion 
is difficult to detect in the true dynamics, since it is 
dominated by the thermal vibrations. 
Presumably this kind of motion is the reason why the inherent trajectory 
in Fig. \ref{fig:Trajectories} shows small displacements 
($\lesssim 0.2\sigma_{AA}$), even when the corresponding true
trajectory seems to oscillate around the same position:
When a transition between inherent structures 
involving significant particle rearrangements in the
surroundings occurs, the particle starts vibrating 
around a position that is slightly displaced, and a corresponding small 
displacement of the inherent trajectory is seen. 
This view of the 
dynamics is also consistent with the fact that the first peak
in the inherent counterpart of $4\pi r^2 G_s(r,t)$ (not shown, 
see Ref. \cite{SchroederDyre}) is not a delta function in $r=0$.
\begin{figure}
\hbox to\hsize{\epsfxsize=1.0\hsize\hfil\epsfbox{
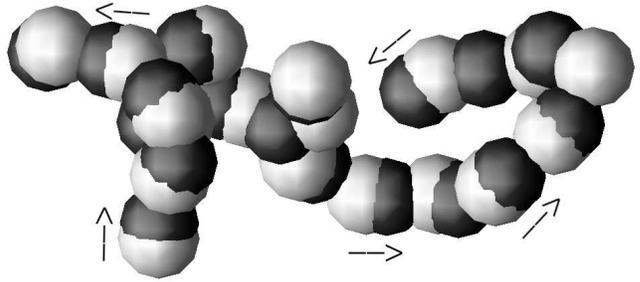}\hfil}
\caption{Before (light) and after (dark) one typical transition, all 
the particles which move a distance greater than 0.2$\sigma_{AA}$. 
Particles are shown 
with a diameter of $\sigma_{AA}$. Note that most particles move 
considerably {\em less} than  $\sigma_{AA}$ 
(compare Fig. \ref{fig:DisplDistr}).
The cooperative, string-like  nature of the particle motions during the 
inter-basin transition can be clearly seen.
}
\label{fig:Transition}
\end{figure}

By observing, for a number of transitions, the positions of all
particles that moved a distance greater than $0.2$ during a transition, we find
these particles to be clustered together in ``strings'', as shown in 
Fig. \ref{fig:Transition}. Typically, one transition appears to involve just 
one string-like cluster. Detailed investigations of the 
transition events will be presented in a separate publication.
Here we simply note that string-like particle motion has been observed
also in the true dynamics above $T_c$ in a similar binary Lennard-Jones 
mixture\cite{ddkppg1}. These strings are found on long time scales and 
involve particles moving approximately one inter-particle distance, 
and are thus different from, but presumably related to, the strings 
found in the present work.

\section{Conclusions}

We have investigated the dynamics
of a model glass-forming liquid in terms of its potential energy landscape 
by  ``quenching'' a time series of MD configurations to a corresponding 
time series of inherent structures.  In this way we have provided
 numerical evidence for the conjecture, 
originally made by Goldstein 30 years ago in this journal 
\cite{Goldstein69}, that below a crossover temperature $T_x$ the 
dynamics of the liquid can be separated into vibrations around and
transitions between inherent structures. Specifically, by comparing the 
self intermediate scattering function $F_s(q,t)$ with  its inherent counterpart
$F_s^I(q,t)$ we presented evidence for the existence of $T_x$. It is 
perhaps not surprising that the dynamics of a liquid becomes dominated 
by the structure of the potential energy landscape at sufficiently 
low temperatures. 
What we have done here using the concept of inherent 
dynamics, is to provide direct numerical evidence for this, 
\emph{and} we have shown that this regime can be reached by 
 equilibrium molecular dynamics (for the particular system 
investigated here). To our knowledge this is the first time such
evidence has been presented.

In agreement with previous proposals\cite{Angell88,Sokolov98,kirk} we 
find $T_x\approx T_c$, where $T_c$ is estimated from a power-law 
fit to $\tau_\alpha$. This is also the temperature range where 
single particle hopping starts to dominate the dynamics, and
 $\tau_\alpha$ becomes on the order of $10^{-9}$ seconds
(Goldstein's estimate of the shear relaxation time at $T_x$). 

The fact that we have been able to cool the system, under equilibrium
conditions, to temperatures where the separation between vibrations
around inherent structures and transitions between these is (almost)
complete, means that it becomes meaningful to study the individual 
transitions over energy barriers, since the transitions in this 
regime dominate the dynamics.
Our two key findings with regards to the individual transitions between 
inherent structures are i)
single particle displacements during transitions show no preference
for displacements on
the order of the inter-particle distance, showing that the single
particle hopping indicated in $4\pi r^2 G_s(r,t)$ at low $T$
(Fig.~\ref{fig:VanHove}) does {\it not} correspond to transitions of
the system over single energy barriers; and ii) particle displacements during
transitions are spatially correlated (in ``strings'').

\section{Acknowledgments}

We thank F. Sciortino and F.H. Stillinger for helpful feedback.
This work was supported in part by the 
Danish Natural Science Research Council.

\end{document}